\documentclass[12pt]{article}
\usepackage{amssymb,amsmath,esint,titlesec}
\titleformat*{\section}{\normalsize\bf}
\titleformat*{\subsection}{\small\bf}

\begin{document}

%%%%%%%%%%%%%%%%%%%%%%%%%%%%%%%%%%%%%%%%%%%%%%%%%%%%%%%%%%%%%%%%%%%%%%%%%%%%%%%

\begin{titlepage}

\setlength{\baselineskip}{18pt}

                               \vspace*{0mm}

                             \begin{center}

{\Large\bf Coarse-graining and symplectic non-squeezing}

                                   \vspace{40mm}

              \large\sf  NIKOLAOS \  KALOGEROPOULOS $^\dagger$\\

                            \vspace{1mm}

  \normalsize\sf  Center for Research and Applications \\
                                  of Nonlinear Systems  \ \  (CRANS),\\
                          University of Patras, Patras 26500, Greece.\\

                            \vspace{2mm}
                         
                                    \end{center}

                            \vspace{20mm}

                     \centerline{\normalsize\bf Abstract}
                     
                           \vspace{3mm}
                     
\normalsize\rm\setlength{\baselineskip}{18pt} 

 \noindent We address aspects of coarse-graining in classical Statistical Physics from the viewpoint 
of the symplectic non-squeezing theorem. We make some comments regarding the implications of the 
symplectic non-squeezing theorem for the BBGKY hierarchy. We also see the cubic cells appearing  in  coarse-graining
as a direct consequence of the uniqueness of Hofer's metric on the group of Hamiltonian diffeomorphisms of  the phase space.    
  
                           \vfill

                                                                         
                             \vfill

\noindent\rule{9cm}{0.3mm}\\  
   \noindent   $^\dagger$ {\normalsize\rm Electronic mail: \ \ \ \  {\sf nikos.physikos@gmail.com}}\\

\end{titlepage}
 
%%%%%%%%%%%%%%%%%%%%%%%%%%%%%%%%%%%%%%%%%%%%%%%%%%%%%%%%%%%%%%%%%%%%%%%%%%%%%%%%

                                                                                \newpage                 

\rm\normalsize
\setlength{\baselineskip}{18pt}

\section{Introduction}

The concept of entropy is fundamental in Statistical Physics. Despite its long history, and its many successes along the way, many of its  aspects are still not very well understood, 
and are sometimes grossly misunderstood,  at both  conceptual and calculational level. This becomes obvious if one even skims the contemporary literature. 
Controversies about entropy are always simmering, periodically resurfacing, but never seem to get settled, or even come close to reaching a consensus. 
Such issues pertinent to entropy and its dual intensive quantity in thermodynamics, namely the temperature, have come up recently 
in the context of conjectured negative absolute temperatures and dark matter, (in-)equivalence of the classical ensembles,
as well as due to the continuing studies of non-equilibrium processes, among many other motivations.
For some relatively recent views on these matters, see for example  \cite{Leb1, Leb2, LY, Sw1, Sw2, DH1, HHD1, SW1, BFS1, SW2, HHD2, Sw3, Sw4, MLGS, BFS2, Sw5, GLTZ}.  \\

In this work, we point out some issues related to the foundational, but still not well-understood issue of coarse graining, from the viewpoint of the symplectic non-squeezing theorem. 
The latter is a fundamental result in symplectic geometry, whose potential significance for Statistical Mechanics has not been fully explored yet. We follow a Hamiltonian  approach ignoring
the role that any metrics on phase space may play, as much as possible. In Section 2, we  present some  well-known facts in symplectic geometry needed in the sequel. In Section 3, we comment on
the relation of the symplectic non-squeezing theorem to the BBGKY hierarchy. In Section 4, we make some general comments about coarse-graining. In Section 5, we present an argument purporting to
explain why cubes are the appropriate cells in which to divide the phase space during  coarse-graining.
Section 6 contains some comments on the relation between the symplectic and the metric structures which may be of  some  interest to Statistical Mechanics and may be the motivation for
 future work in the metric approach to determining the statistical properties of systems of many degrees of freedom.\\ 

 %%%%%%%%%%%%%%%%%%%%%%%%%%%%%%%%%%%%%%%%%%%%%%%%%%%
 
 \section{Symplectic Preliminaries}
 
Let us consider a classical particle system having \ $n$ \  degrees of freedom  in  3-dimensional space. Such a system is usually analyzed either in a Lagrangian or in a Hamiltonian way.
The Hamiltonian approach  has the advantage over the Lagrangian one that it allows the use of additional symmetries existing between the canonical positions and momenta  which may 
not be manifest in the Lagrangian description.  Such symmetries are symmetries of the underlying system in phase space and are central in determining aspects of coarse graining, 
as will be elaborated upon in the sequel.  Hence we proceed in the present work following the Hamiltonian approach.  
We use \cite{McDS} as our basic reference for symplectic geometry in the sequel. 
Let the microscopic Hamiltonian of the system under study be denoted by \ $\mathcal{H}$ \ 
and the phase space \ $\mathcal{M}$ \  be parametrized by  canonical coordinates 
\ $(q_i, p_i), \ i=1, \ldots, 3n$. \  The phase space \ $\mathcal{M}$ \ is naturally endowed with a symplectic structure \ $\omega_0$ \ which is locally expressed, according to Darboux's theorem, as 
\begin{equation}
           \omega_0 = \sum_{i=1} ^{3n} dq_i \wedge dp_i              
\end{equation}
Symplectic manifolds such as \  $(\mathcal{M}, \omega_0)$ \ are devoid of any local geometric structure. Indeed all such symplectic manifolds of the same dimension are locally diffeomorphic to \  
$(\mathbb{R}^{6n}, \omega_0)$. \ 
This statement is  intimately related to the fact that the group of symplectic diffeomorphisms/canonical transformations is infinite dimensional, 
in sharp contrast to the group of isometries of a Riemannian manifold. As a result, all symplectic manifolds of the same dimension are locally ``the same". 
As long as they are of the same dimension, such manifolds only differ from each other in their global aspects, which are encoded in their  topology.  
For this reason, the existence of the canonical symplectic structure \ $\omega_0$ \ alone on \ $\mathcal{M}$ \ 
may be considered to be too ``rigid" for an effective description of the behavior of the  associated physical system. 
After all, the overwhelming majority of physical theories do have local degrees of freedom which distinguish  one from another. \\

If we want to remain within the realm of symplectic manifolds for Hamiltonian systems, 
without assuming an additional structure on the phase space \ $\mathcal{M}$ \ such as a metric, or some conditions such as integrability of the associated almost complex structure, 
then we can proceed as follows. We know that Hamiltonian flows preserve the symplectic volume: this is Liouville's theorem. Its immediate result is that the Gibbs entropy of an isolated system remains constant 
if the system is  described by a Hamiltonian flow. But, symplectic maps turned out to have a more fundamental property which distinguishes them from volume-preserving maps: the symplectic non-squeezing property 
and its generalizations. There are symplectic invariants called ``symplectic capacities" \cite{McDS, HoferZ} which, incidentally, have nothing to do with the volume of the underlying manifold, which 
volume-preserving flows do not possess. 
The non-squeezing property essentially quantifies that the symplectic maps are more ``rigid" than the volume-preserving ones, and that the former are a proper subset of the latter.  To be more precise, it 
establishes that the set of symplectic diffeomorphisms of \ $\mathbb{R}^{2n}$ \ is \ $C^0$-closed in the set of all diffeomorphisms of \ $\mathbb{R}^{2n}$, \ rather than its closure being the set of volume-preserving 
diffeomorphisms of \ $\mathbb{R}^{2n}$ \ as one might possibly have expected. \\

The symplectic non-squeezing theorem was first proved by M. Gromov in \cite{G-Pseudo}, subsequent proofs are given by \cite{EkelHof, Vit, HoferZ}, and since that time it has been the foundation of an 
explosive growth in the field of symplectic geometry.  To formulate it, let  \ $\| x \|$ \ indicate the Euclidean norm of \ $x\in \mathbb{R}^{2n}$ \  and let  
\begin{equation}
      B_{2n}(r) = \left\{ x\in \mathbb{R}^{2n}: \ \| x \| \leq r \right\}
\end{equation}
denote the closed ball of radius \ $r$ \ centered at \ $0\in\mathbb{R}^{2n}$. Let 
\begin{equation}
       Z_{2n}(r) = B_2(r) \times \mathbb{R}^{2n-2} = \left\{ x\in \mathbb{R}^{2n}: q_i^2 + (p_1)^2 \leq r^2\right\}
\end{equation}
be the cylinder in \ $\mathbb{R}^{2n}$ \ whose base lies in the symplectic 2-plane \ $(q_1, p_1)$ \ having base radius \ $r$. \ If there exists a symplectic embedding \ $B_{2n}(r) \hookrightarrow Z_{2n}(R)$, \ 
then \ $r\leq R$. \  This ``non-squeezing theorem"  is not valid if the base of the cylinder lies on an isotropic 2-plane, such as \ $(q_1, q_2)$.  \ The theorem was extended by F. Lalonde and D. McDuff \cite{LMcD} to any
symplectic manifold.  During the last two decades, several more rigidity results have been obtained for symplectic embeddings \cite{Schlenk}, of mostly  2- or 4-dimensional symplectic objects having a simple geometry 
onto similar classes of objects or their stabilized analogues. From the viewpoint of Statistical Mechanics, it would be desirable to have similar results on embeddings of high dimensional symplectic manifolds, 
which are not stabilized versions of lower dimensional ones. However, no such results of potential physical interest have been obtained, so far as the author knows. \\

The symplectic non-squeezing theorem can be reformulated  as follows \cite{EliashG}. Let \ $\mathbb{P}$ \ denote the orthogonal projection, with respect to the Euclidean metric, of an object in \ $\mathbb{R}^{2n}$ \ 
onto the 2-dimensional symplectic plane \ $(q_1, p_1)$ \ and let \ ${\sf Vol}_2$ \   indicate the symplectic volume (``area") of a 2-dimensional set on the \ $(q_i, p_i), \ i=1,\ldots, 3n$ \  plane. 
Consider a symplectic embedding \ $\varphi: B_{2n}(r)\hookrightarrow \mathbb{R}^{2n}$. \ 
Then  
\begin{equation}
        {\sf Vol}_2 (\mathbb{P}\varphi (B_{2n}(r)) \geq \pi r^2
\end{equation}
Moreover, it is known \cite{AbbMat} that analogues of (4) exist for intermediate dimensions, namely that 
\begin{equation}
        {\sf Vol}_{2k}(\mathbb{P}\varphi (B_{2n}(r))) \geq \gamma_{2k} r^{2k},    \hspace{15mm} 1\leq k \leq n
\end{equation} 
but only for linear symplectic automorphisms \ $\varphi: \mathbb{R}^{2n} \rightarrow \mathbb{R}^{2n}$ \ onto complex linear subspaces of \ $\mathbb{R}^{2n}$ \ of real dimension \ $2k$. \ In  (5), 
\ $\gamma_{2k}$ \ stands for the volume of the $2k$-dimensional ball \ $B_{2k}(1)$ \ of unit radius.
Such a result does not hold for general, 
non-linear, symplectic maps \cite{Guth, AbbMat}.  Indeed, for every \ $\varepsilon >0$ \ there exists a smooth symplectic embedding \ $\varphi: B_{2n}(1) \rightarrow \mathbb{R}^{2n}$ \ such that 
\begin{equation}
       {\sf Vol}_{2k}(\mathbb{P}\varphi (B_{2n}(1))) < \varepsilon
\end{equation}  
Hence, the intermediate dimensional non-squeezing theorem does not hold anymore when passing from linear to non-linear symplectic maps \ $\varphi$. \ The transition from linear to non-linear symplectic maps 
resulting in the transition from the validity to the non-validity of the symplectic non-squeezing theorem for intermediate dimensions was examined in  detail in \cite{AbbMat}.  \\

The conclusion of the symplectic non-squeezing theorem does not hold, either for linear or non-linear symplectic maps,  if the word ``projections" is replaced by the word ``sections" in the above statements. 
Due to (5), (6), we only have to formulate the case for sections for \ $\varphi$ \ being a linear symplectic map. Let \ $\mathcal{V} \subset \mathbb{R}^{2n}$ \ be a complex linear subspace of real dimension 
\ $2k, \ 1\leq k \leq n$.  \ Then
\begin{equation}
       {\sf Vol}_{2k} (\mathcal{V} \cap \varphi (B_{2n}(r))) \leq \gamma_{2k} r^{2k}
\end{equation}
and the equality holds if and only if the linear subspace \ $\varphi^{-1}(\mathcal{V})$ \  is complex. \\

Despite its foundational role in geometry, the symplectic non-squeezing theorem, further developments stemming from it, and its applications in Mathematics, have been largely ignored by the Physics community. 
An exception to this is the string theory community where the ideas contained in \cite{G-Pseudo}, which gave rise to the various versions of Floer homology and quantum cohomology, are used in a variety of contexts, 
not the least of which are the various statements associated with mirror symmetry \cite{Mirror}. Another  exception has been the work of M. de Gosson who has strongly advocated for the use of the symplectic 
non-squeezing theorem in Physics in general, but especially in the context of the semi-classical limits of quantum systems \cite{DeG1, DeG2, DeG3, DeG4, DeG5, DeG6, DeG7, DeG8, DeG9, DeG10} 
during the last two decades.  Being influenced by his work, we also presented some views on the possible implications of the non-squeezing theorem for the macroscopic time irreversibility of 
Hamiltonian systems in \cite{Kal1, CrKal}.  These contributions illustrate the viewpoint which is adopted and has largely motivated the present work. \\  

%%%%%%%%%%%%%%%%%%%%%%%%%%%%%%%%%%%%%%%%%%%%%%%%%%%%%%%%%%%%%%%%%%%%%%%%

\section{Symplectic non-squeezing and the BBGKY hierarchy}

One could argue that for a systems with many degrees of freedom which are modelled by  flows on high-dimensional manifolds like \ $\mathcal{M}$, \ 
statements like (4) which involve projections onto 2-dimensional symplectic planes cannot really be of any significance. In other words, for a Hamiltonian evolution in a  high dimensional phase space 
\ $\mathcal{M}$ \ (4) is too weak of a constraint, or provides too small a ``rigidity" due to its dimension alone,  to be of any significance in determining the macroscopic properties of the underlying system. \\

One could start uttering some objections this viewpoint, which is very reasonable, by invoking the following argument. As is well known, the behavior of a set can be determined either directly, geometrically, or 
by determining the behavior of appropriate sets of functions defined on it. The most obvious such choice is the characteristic function of the set. Micro-canonical probability distributions of isolated systems 
are uniform on the sets of constant energy, therefore they are constant multiples of such characteristic functions. So, instead of working directly with sets, one could as well work with the micro-canonical probability 
densities. \\

In this spirit, consider dual description of a phase space and its subsets, in terms of probability distributions defined on them. The Bogoliubov-Born-Green-Kirkwood-Yvon (BBGKY) hierarchy 
which is extensively used in the derivation of kinetic equations for particle systems  describes the time evolution of the marginal distribution functions of such particle systems, starting from Liouville's equation 
for the distribution function of all particles in the system  and gradually descending to marginals of fewer and fewer particles until it reaches the distribution function of a single particle \cite{CDFR}. 
To be more concrete, let \ $f_n (q_1, p_1; q_2, p_2;  \ldots ; q_n, p_n; t)$ \ denote the joint probability density function describing the behavior of all  \ $n$ \ particles in the system under study. Since 
the system is not assumed to be in equilibrium, then \ $f_n$ \ can explicitly depend on some variable \ $t$ \ describing evolution, let us call it ``time".  Since \ $f_n$ \ is a probability density it is 
normalized as 
\begin{equation}
      \int_\mathcal{M} \ f_n(q_1, p_1; q_2, p_2; \ldots; q_n, p_n; t) \  d{\sf Vol}_{6n} 
\end{equation}  
where the volume of the phase space \  $\mathcal{M}$ \  is
\begin{equation}
     d{\sf Vol}_{6n} = \prod_{i=1}^n \  d^3 q_i \  d^3p_i
\end{equation}
For most macroscopic systems \ $n\sim 10^{23}$. \ Ideally, we would like to be able to determine \ $f_n$ \ since knowing it would allow us to calculate any quantity of physical interest related to the system,
as well as the evolution of any such quantity. Of course this is not possible practically, nor is necessarily required for an effective macroscopic description of the system, so we have to, and we can,  settle with less.  
Liouville's theorem is the statement that 
\begin{equation}
        \frac{df_n}{dt} = 0
\end{equation}
namely that the co-moving volume of a system on \ $\mathcal{M}$ \  remains invariant under its Hamiltonian evolution. This theorem does not state anything, however,  about the evolution of the exact shape of 
this volume, something which is left to be found as a solution of Hamilton's equations, or equivalently, it is determined by the evolution generated by the Hamiltonian vector field \ $X_{\mathcal{H}}$. \ 
Such a vector field \  $X_\mathcal{H}$ \  associated to the Hamiltonian function \ $\mathcal{H}$ \  is defined by requiring it to obey
\begin{equation}
                  \omega_0(X_\mathcal{H}, \cdot) = d\mathcal{H}
\end{equation}

The Poisson bracket between two vector fields \ $X_\mathcal{A}, \ X_\mathcal{B}$ \ associated to the scalar functions \ $\mathcal{A}, \ \mathcal{B}: \mathcal{M} \rightarrow \mathbb{R}$  \ respectively, 
is defined by
\begin{equation}
            \left\{\mathcal{A}, \mathcal{B} \right\} = \omega_0 (X_\mathcal{A}, X_\mathcal{B})  
\end{equation}
The one-particle probability density is defined as 
\begin{equation}
     f_1 (q_1, p_1;t)  \ =  \ n \int_{\mathcal{V}_{n-1}} \  f_n(q_1, p_1; q_2, p_2; \ldots; q_n, p_n; t) \  \prod_{i=2}^n \ d^3q_i \ d^3p_i
\end{equation}
which is normalized as
\begin{equation}
       \int_{\mathcal{V}_1} \ f_1(q_1, p_1; t) \ d^3q_1 \ d^3p_1 \ = \  n
\end{equation} 
with \ $\mathcal{V}_{n-1}$ \ being the submanifold of \ $\mathcal{M}$ \ which is parametrized by \ $(q_2, p_2; \ldots; q_n, p_n)$ \ and 
$\mathcal{V}_1$ \ being the submanifold of \ $\mathcal{M}$ \  which is parametrized by \ $(q_1, p_1)$. \ In a similar manner, and for identical particles, one defines the 2-particle distribution function by 
\begin{equation}
   f_2 (q_1, p_1; q_2, p_2; t) \ = \ n(n-1) \int_{\mathcal{V}_{n-2}} \ f_n(q_1, p_1; q_2, p_2; \ldots; q_n, p_n; t) \  \prod_{i=3}^n \ d^3q_i \ d^3p_i
\end{equation}
and, more generally, the marginal $k$-particle distribution function for \ $1\leq k \leq n$ \ is defined as 
\begin{equation}
   f_k (q_1, p_1; q_2, p_2, \ldots; q_k, p_k; t) \ = \ \frac{n !}{(n-k) !} \int_{V_{n-k}} \ f_n(q_1, p_1; q_2, p_2; \ldots; q_n, p_n; t) \  \prod_{i=k}^n \ d^3q_i \ d^3p_i
\end{equation}

Let us assume that the Hamiltonian of the particle system is  
\begin{equation}
              \mathcal{H} \ = \ \sum_{i=1}^n \frac{p_i^2}{2m_i} + \sum_{i<j} U(q_i-q_j)  
\end{equation}
where \ $U$ \ stands for the potential energy between pairs of particles. We ignore  contributions to \ $U$  \ from triplets and higher numbers of particles.
 Then the evolution of the marginal distribution \ $f_1$ \ is given by 
\begin{equation}
     \frac{\partial f_1 (q_1,p_1;t)}{\partial t} \  = \  \left\{\mathcal{H}_1, f_1 \right\} + \int \frac{\partial U(q_1-q_2)}{\partial q_1} \cdot \frac{\partial f_2}{\partial p_1}  \ d^3q_2 \ d^3p_2
\end{equation}
where
\begin{equation}
            \mathcal{H}_1 = \frac{p_1^2}{2m_1}
\end{equation}
and \ $\cdot$ \ indicates the inner product on \ $\mathcal{M}$. \  In general
\begin{equation}
     \frac{\partial f_k}{\partial t} \ = \ \left\{ \mathcal{H}_k, f_k \right\} + \sum_{i=1}^k \int \frac{\partial U(q_i-q_{k+1})}{\partial q_i} \cdot \frac{\partial f_{k+1}}{\partial p_i}  \ d^3q_{k+1} \ d^3p_{k+1}
\end{equation}
where 
\begin{equation}
       \mathcal{H}_k \ =  \ \sum_{i=1}^k \frac{p_i^2}{2m_i} + \sum_{i<j\leq k} U(q_i-q_j) 
\end{equation}

As is well-known and can be immediately seen, the BBGKY hierarchy (20), (21) is of little practical interest unless one can find a plausible set of arguments allowing them to truncate it at the distribution 
function \ $f_k$ \ of a very small number of particles, such as \ $k=3$ \ for instance.  
As a result, in the BBGKY hierarchy a central role is played by the $1-$ and the $2-$ particle distribution functions \ $f_1$ \ and \ $f_2$ \ given by (13) and (15) respectively.  
Hence the most important distribution functions one is dealing with in the BBGKY hierarchy have support on $2-$ and $4-$ dimensional subsets of the phase space \ $\mathcal{M}$. \ 
For this reason,  additional constraints placed upon the $1-$ and $2-$ particle distribution functions which are not related to Liouville's theorem may have substantial consequences 
for the description of the evolution the whole system. The symplectic non-squeezing  theorem (4) provides such a constraint for the $1-$particle distribution function, as its validity is independent of the 
validity of Liouville's theorem. This is true since in the  probabilistic  language, the $1-$particle distribution functions are marginals, therefore they correspond to the projections of  their support to the 
$2-$dimensional symplectic planes of the canonically conjugate variables.\\ 

 Consider a system with many degrees of freedom whose initial probability distribution has as support a Euclidean  ball of dimension \ $6n$ \ and radius \ $R$. \ The physical model that we have in mind is an isolated 
 system of coupled mechanical harmonic oscillators of equal masses and equal  frequencies. The total energy of the system is the only constraint hence it gives rise to this ball in \ $\mathcal{M}$. \
 Granted that this is an integrable system, so it is very unlike the typical systems one encounters in Statistical Mechanics. Still, we 
 use it for illustrative purposes. In the evolution of such a system, the support of the probability distribution will change shape but will keep its overall 
 volume constant. This is described by stating that \ $f_{n}$ \ should be a constant. However, according to the symplectic non-squeezing theorem this ``micro-canonical" distribution is not the only constraint that \ $f_n$ \ 
 should obey. Each marginal distribution \ $f_1$ \ should also have as support a disk whose radius is at least \ $R$. \ As the system evolves, in order to keep its overall symplectic volume fixed, it will have to change shape, as is 
 dictated by the Hamiltonian flow. For a small amount of time though, the shape of such a support will not change appreciably from that of a ball, if the ball is relatively small. 
 Its projections to symplectic 2-planes will be almost spherical but will have a radius that will be at least as big as \ $R$. \ Hence the value of \ $f_1$ \ will decrease on its support as a function of time. \\

 %%%%%%%%%%%%%%%%%%%%%%%%%%%%%%%%%%%%%%%%%%%%%%%%%%%%%%%%%%%%%%%%
 
 \section{On coarse-graining}
 
 The issue of coarse-graining is fundamental in Statistical Physics. It expresses the fact that in order to determine the macroscopic behavior of a system, many of its microscopic features have to be 
 ignored. At the purely classical level, there is no obvious  way to implement such a coarse graining. One usually considers the phase space \ $\mathcal{M}$ \  and partitions it in small cubes. The only requirement 
 is that such cubes should be much bigger in size than any of the microscopic details of the Hamiltonian system describing the physical system under study, so one can perform some kind of averaging inside
 each such a cube and in essence ``gloss over"  the unimportant features of each of the microscopic degrees of freedom of the system. \\
 
 The view of Boltzmann is to enumerate all such microscopic states, let their number  be denoted by \ $\mathcal{N}$, \ giving rise to the same
 macroscopic state in determining the entropy of the system, whether it is in equilibrium or not. 
 Let the coarse-grained phase space of the system be denoted by \ $\overline{\mathcal{M}}$. \ 
 The overwhelming majority of the microscopic states of \ $\overline{\mathcal{M}}$ \ correspond to macroscopic equilibrium, 
 which justifies the increase of the Boltzmann entropy
 \begin{equation}
        \mathcal{S}_B = k_B \log \mathcal{N} 
 \end{equation}
 for an isolated  system approaching equilibrium, where \ $k_B$ \  is Boltzmann's constant.\\
 
 The view of Gibbs, who instead works with probability distributions in his ensemble theory, is somewhat dual:
  one replaces the given probability distribution \ $\rho$ \ on phase space \ $\mathcal{M}$, \ with its (constant)  average value \ $\overline{\rho}$ \ inside each cube/element of \ $\overline{\mathcal{M}}$ \ and 
  then calculates the coarse-grained Gibbs entropy 
  \begin{equation}
                  \overline{\mathcal{S}_G}[\overline{\rho}(x)] \ = \  - k_B \int_{\overline{\mathcal{M}}} \ \overline{\rho}(x) \ \log{\overline{\rho}(x)} \ d\mathsf{Vol}_{\overline{\mathcal{M}}}   
  \end{equation} 
  In (23) one should be, strictly speaking, using a summation instead of an integration, but due to the small sizes of the cubes of the partition from a macroscopic viewpoint, 
  integration works in practice equally well.
 Then, one sees that even though the Gibbs entropy expressed through the actual probability distribution \ $\rho$ \ cannot change for an isolated system due to Liouville's theorem, 
 its coarse-grained counterpart \ $\overline{\mathcal{S}_G}$ \ does change as the system approaches equilibrium.\\
 
 The existing literature on coarse-graining is vast, because such a central issue has 
 occupied the attention of practitioners since its earliest days, when it was introduced by Gibbs, and by P. and T. Ehrenfest \cite{EE}. 
 From the time of its introduction, all kinds of works have appeared covering aspects of coarse-graining, ranging from the purely philosophical/ontological to the very formal ones. 
 Some indicative recent references are \cite{Leb1, Leb2, GLTZ, GMH, Frigg, Gorban, CFLV, ASV, TeSte, SAD}. In general, most works referring to the foundations of Statistical Mechanics 
 and the meaning of entropy, have some discussion on coarse-graining.\\
 
  A troubling aspect of this process, is that even though there is considerable arbitrariness in performing a coarse-graining, 
 the final results arising from calculations of the resulting thermodynamic potentials are physical/measurable quantities, hence completely independent of any subjective choices leading to such a coarse-graining.
 This may be due to statistics, where features of individual variables are glossed over in favor of statistically significant quantities. As such, reasonable but subjective choices leading to coarse-graining of particular systems 
 may be statistically unimportant. Hence  statistics seems to effectively eliminate subjectivity. \\
 
   The question that we address in this section is whether there is any preferred, or even optimal, way of performing such a coarse-graining. A related question is what, if any, constraints should be placed to such a 
 coarse-graining by the microscopic features of the system. At a first thought, the answer is negative to both questions. Coarse-graining is, after all, a process that someone performs ``by hand" when making a 
 transition from the microscopic to a mesoscopic or macroscopic description of a system and there is a substantial degree of arbitrariness to it. However, this question is crucial, since the calculation of entropy
 or any of the associated thermodynamic potentials relies on coarse-graining. \\
 
 If one adopts the viewpoint of Boltzmann the description of a macroscopic system, then there is no obvious way of choosing how to coarse-grain the phase space \ $\mathcal{M}$. \  To get around this difficulty, one
 uses inherent uncertainties about a physical system:  it is actually impossible to measure with absolute accuracy the properties of any physical system, even in classical Physics. Hence uncertainties can creep in two
 places. There are uncertainties in the Hamiltonian of the system; indeed, no system can be absolutely isolated, except the Universe itself. This lack of absolute isolation can be described either by introducing terms coupling the 
 system under study to its environment, or by introducing the possibility for fluctuations of the degrees of freedom in the underlying Hamiltonian. It is not unreasonable to assume that such fluctuations are small, certainly away from 
 phase transitions, and therefore treat them as small perturbations of the underlying Hamiltonian. Incidentally, these two approaches give equivalent results, heuristically 
 at least, in the perturbative regime of relativistic quantum field theories, which are systems however that we do not discuss in the present work. \\
 
 These small perturbations can either be described either 
 as being stochastic  (``noise") or as being sustained for particular systems. As a result, the evolution of the Hamiltonian system is not described by a single trajectory in \ $\mathcal{M}$ \  but by the evolution 
 of a small volume around it, which expands at some rate depending on the perturbation, as the system evolves. This is the process of  $\varepsilon$-fattening of the trajectories of the underlying Hamiltonian system. 
 The fact that this is the generic behavior, rather than the exception for multi-dimensional dynamical systems, was proved in \cite{Smale}. 
 Given this, someone can coarse-grain \ $\mathcal{M}$ \ by making a judicious choice of scale based on the features of such a perturbation and the subsequent rate of volume expansion. The process of $\varepsilon$-fattening
 can be most effectively described in a Riemannian or metric-measure theoretical setting, so we will not elaborate on it any further.  \\
      
 A second way to introduce coarse-graining into the system is by considering not just one, but a whole set of initial conditions. This reflects our lack of absolute precision in preparing the physical system. Such a set of initial 
 conditions can have any ``shape" in phase space, but has  to be ``small" in order to reflect the small macroscopic uncertainties in preparing the system. For simplicity, which however is compatible with the  symplectic structure as 
 will be explained in the sequel, let us choose such a set of initial conditions to have the shape of a ball around the desired initial conditions of the system. Such a ball provides a natural coarse-graining, a natural lower 
 cut-off scale in \ $\mathcal{M}$. \  The fact that its volume remains invariant under the Hamiltonian flow is guaranteed by Liouville's theorem. We also know from the symplectic non-squeezing theorem that its projections    
 onto the symplectic $2-$planes of the conjugate variables will also have to remain invariant or increase in size as the system evolves. There is no required minimum size of such initial conditions in classical physics; 
 in fact it is exactly zero.   For further discussion, see \cite{DeG1, DeG2, DeG3, DeG5, DeG6, DeG8} .The scale for  such a minimum size is set by Quantum Physics. 
 This approach, in essence, allows us to speak about a minimal size, as for the case of ``quantum blobs" \cite{DeG5}.\\
 
 However, we know of no constraints on  the projections of the initial conditions  on the isotropic  $2-$planes. Hence, the symplectic embeddings give no information on the geometric features of $2-$dimensional isotropic 
 submanifolds. This is not totally surprising $2-$dimensional symplectic planes are essentially almost complex curves embedded in a very high dimensional symplectic manifold. 
 It seems unlikely that the geometry of such curves may have anything to say about the 
 ambient space. The situation is markedly different for Lagrangian subspaces though, since the latter are maximal dimensional isotopic sub-manifolds of the ambient space, 
 which allows them to have a special significance in symplectic geometry \cite{McDS}. It is not an accident that pseudo-holomorphic curves \cite{G-Pseudo}, which are objects having real dimension $2$, 
 have been applied most effectively in symplectic and enumeration problems on $4-$dimensional manifolds.  From our viewpoint, it may be worth pointing out, that Lagrangian sub-manifolds inside any constant 
 energy hyper-surface of a symplectic manifold, as we assume in our case, remains invariant under the Hamiltonian flow. This fact might have considerable consequences if a middle-dimensional symplectic 
 non-squeezing theorem were valid which, as \cite{AbbMat} argued, does not, except for the case of linear symplectic maps.\\          
 
  %%%%%%%%%%%%%%%%%%%%%%%%%%%%%%%%%%%%%%%%%%%%%%%%%%%%%%%%%%%%%
  
 \section{On the shape of the coarse-graining cells}

In Section 2, we discussed the case of symplectic embeddings of balls and its implication for the marginal ($1-$particle) distributions appearing in the BBGKY hierarchy. 
Such balls, even if convenient objects from a metric viewpoint, appear not have the appropriate shape  for providing a coarse-graining of the phase space \ $\mathcal{M}$ \ of the system. 
As was previously pointed out, such coarse-graining usually takes place by using cubes/parallelepipeds whose sides are parallel to the coordinate axes, at least as long as \ $\mathcal{M}$ \ is \ $\mathbb{R}^{6n}$. \ 
From a metric viewpoint, one could not easily expect that a cube and a  ball are close to each other. In a sense, these are two convex bodies that are as far from each other as possible, at least when one uses 
the Banach-Mazur distance \ \cite{MilS}. \  We have used this maximal distance/extremal mismatch to argue about coarse-graining in \cite{Kal3}.   \\

In a symplectic context things are radically different though. To explain this, let us consider instead of \ $\mathcal{M}$ \ which is usually non-compact, a compact approximation. This is quite common in
mainstream treatments, ones not aspiring to mathematical rigor of physical systems, at least in position space: we resort to systems confined in a cubic box, with appropriate boundary conditions. We then let the  
width of such a box increase without any upper bound. Similar things can be done about canonical momenta, as in hydrodynamics, when one provides an upper bound to the magnitude of the  momenta of the 
particles of the system by confining our attention to the behavior of the low-frequency/large wavelength modes. To proceed and to be in accordance with the fact that cubes  are used in  typical coarse grainings of \
$\mathcal{M}$, \  let us investigate how close is such a cube to balls of equal dimension \ $2d$, \  from the viewpoint of symplectic maps. \\

We can rephrase this question as a packing problem: consider the unit
cube \ $C_{2d} = [0,1]^{2d}$ \ in \ $\mathbb{R}^{2d}$. \ What fraction of the volume of \  $C_{2d}$ \  can someone fill  by using symplectic embeddings of \ $k$ \  equal disjoint balls? If \ $k$ \ is large, the answer is: \ $1$. \  
To be more precise, let \ $\mathcal{P}_k (\mathcal{M}, \omega )$ denote the $k$-th packing number of a compact $2d$-dimensional symplectic manifold \ $(\mathcal{M}, \omega)$. \ This is defined as 
\begin{equation} 
                  \mathcal{P}_k (\mathcal{M}, \omega ) = \frac{ \sup_r \mathsf{Vol}_{2d} (\sqcup_k B_{2d}(r))}{\mathsf{Vol}_{2d} (\mathcal{M}, \omega)}
\end{equation}
where the supremum is taken over all radii \ $r$ \ for which there exists a symplectic embedding of the disjoint union of \ $k$ \ balls of radius \ $r$ \ into \ $(\mathcal{M}, \omega)$.\  
Then \ $(\mathcal{M}, \omega)$ \ 
has packing stability if there exists some integer \ $n_0 (\mathcal{M}, \omega)$ \ such that  \ $\mathcal{P}_l (\mathcal{M}, \omega) = 1$ \ for all \ $l \geq n_0 (\mathcal{M}, \omega)$. \ In \cite{BuseH} it was proved that 
if \ $(\mathcal{M}, \omega)$ \ is rational, namely if \ $[\omega] \in H^2(\mathcal{M}, \mathbb{Q}) \subset H^2(\mathcal{M}, \mathbb{R})$, \  then it has packing stability. 
The  condition on the second singular homology group of \ $(\mathfrak{M}, \omega)$ \ is obviously satisfied for the cube \ $C_{2d}$, \ since the cube is homologically trivial. \ 
The same result applies if one uses small ellipsoids instead of balls \cite{BuseH}. \ As a result of this theorem, one can use sufficiently many balls \ $(k\rightarrow\infty)$, \  instead of cubes,  for coarse graining the phase 
space \ $\mathcal{M}$ \  of a physical system of many degrees of freedom. \\

A  question that naturally arises has to do with the shape of the allowed partitions of phase space. It was pointed out initially by Boltzmann, and ever since, that an arbitrary partition of phase space  is not appropriate if one wishes to 
get correct physical results. 
Clearly cubes or parallelepipeds are the first obvious choice based on naturality and simplicity. However, the experience of integration theory makes us skeptical of adopting such ``obvious" choices without further investigation. 
We would expect any partition of the phase space to be acceptable, if we accept that partitions are man-made devices allowing us to proceed with our coarse-graining arguments and beyond. However this is obviously not true.
Therefore, there may be more to phase space coarse-graining cells than just being a convenient device arbitrarily chosen ``by hand" to serve our purposes.\\

It turns out that to get correct results on has to use a ``right" or the ``right" partition.  It seems that phase space partition through cubes whose sides are parallel to the local canonical position  and 
momenta axes gives correct results. Some other partitions do not. But it is unclear why a partition by such cubes works, or what other partitions might  work and how to distinguish good/acceptable partitions 
from not good ones (see \cite{Frigg} pp. 27-28). \\

A brief digression is in order here. 
The question of cubes as fundamental cells of coarse-graining in phase space  is strongly reminiscent of  a similar question in Quantum Mechanics: 
the uncertainty principle of Quantum Mechanics has its simple commutator form only if one uses Cartesian coordinates. For any other coordinate system,
 the canonical position and momentum commutator is not so trivial,  or natural, to obtain. 
 However, it is widely accepted in Physics, at least since the advent of General Relativity that coordinates  have no fundamental meaning in a physical theory, 
 but are just labels associated to a particular observer. They may be convenient technical devices allowing someone to perform explicit calculations in a specific context, 
but the results of physical measurements should either not depend on them at all, or might depend on them in a ``covariant" way. 
The question of why the Cartesian coordinates seem to play this special role in the formulation of the uncertainty principle has been 
around since the earliest days of Quantum Mechanics and but has not been adequately resolved. The advent of  geometric quantization did little to answer such a question, apart from 
placing it in a manifestly coordinate independent form. Maybe an answer to this question could be found along the lines that we propose for finding a resolution to the issue of coarse-graining of 
phase spaces in this section.  After all, the probabilistic philosophy and formalism established in Classical Statistical Mechanics pervades Quantum Mechanics, despite their different domains of applicability, 
goals and formalisms. Maybe the analogies that may be drawn, and the lessons learned, could be transferred from one of these disciplines to the other.\\ 

In what follows, we attempt to provide a possible answer to the  question regarding the ``right" shape(s) of the partition cells of phase space by resorting to dynamics. 
One could  argue about using such cubes for the phase coarse-graining, based on simplicity, or even elegance, as noted above. 
But such an approach is not convincing enough. Something stronger is needed. Aesthetic or simplicity considerations do not really shed any light into the suitability of such 
cubes as fundamental cells of the phase space coarse-graining. \\

To achieve our goal we will be examining the geometry of the set of Hamiltonian diffeomorphisms on the given phase space \ $(\mathcal{M}, \omega)$ \cite{Polt}.  
We see, to begin with, that a cube with sides parallel to the coordinate  axes in \ $\mathbb{R}^{n}$ \ is the unit ball of the \ $L^\infty$, \ otherwise known as the $\sup$-norm.  
This is  defined as follows. Let \ $f: \mathbb{R}^n \rightarrow \mathbb{R}$ \ be a measurable function. 
Define the Lebesgue spaces  \ $L^s (\mathbb{R}^n, \mathbb{R})$ \ as the set of such functions so that their $s$-norm is finite
\begin{equation}
            \| f \|_s =  \left\{ \int_{\mathbb{R}^n} |f|^s \  d\mathsf{Vol}_{n}  \right\} ^\frac{1}{s} \ < \ \infty \hspace{20mm} 1\leq s < \infty
\end{equation}
and, in addition, define the \ $L^{\infty}(\mathbb{R}^n, \mathbb{R})$ as the Lebesgue space whose $\sup$-norm is finite  
\begin{equation}
      \| f \|_\infty = \sup \left\{ |f(x)|: \ x\in \mathbb{R}^n \right\} < \infty
\end{equation}
It is not too far-fetched to expect that 
\begin{equation}
        \lim_{s\rightarrow \infty} \|f\|_s = \| f \|_\infty   
\end{equation}
which turns out to be true. These definitions can be straightforwardly extended to $n$-dimensional differentiable manifolds \ $\mathcal{M}$.  \\

Let \  $(\mathcal{M}, \omega)$ \ be a closed symplectic manifold. Most of the times in Physics, the phase space is a cotangent bundle, therefore non-compact. 
However, we can use compact approximations as regulators, whose diameters and widths  at the end of each calculation are sent to infinity, as we commented above. 
This approach can be dually expressed by considering only compactly supported sets of functions on phase space.\\ 

To simplify matters, we will also assume that $\mathcal{M}$ \ has no boundary. 
We could use the set of all smooth functions on \ $\mathcal{M}$, \ but since it is compact, we can equivalently use the set of all mean zero value functions on it, which we denote 
by \ $C_0 ^\infty(\mathcal{M})$. \ The results of this and the next paragraphs hold for general,  non-autonomous Hamiltonians \ $\mathcal{H}_t$ \ and their corresponding 
Hamiltonian vector fields \ $X_{\mathcal{H}_t}$ \ even though that level of generality is not needed in this work.  
Consider the group of Hamiltonian diffeomorphisms \ $\mathsf{Ham}(\mathcal{M}, \omega)$ \ of such a phase space \ $\mathcal{M}$ \
which consists of all Hamiltonian flows whose time parameter has been set to \ $1$.  \ This is an infinite dimensional Lie group. The corresponding Lie algebra \ $\mathcal{G}$ \  is
 \ $C_0^{\infty} (\mathcal{M})$. \  Let  \ $\| \cdot \|$ \  be a norm on \ $\mathcal{G}$. \ We can use \ $\| \cdot \|$ \ to define the length of the flow \ 
 $\phi: [0, 1] \rightarrow \mathsf{Ham}(\mathcal{M}, \omega)$ \ by   
\begin{equation}
              l(\phi) \ = \  \int_0 ^1 \| \frac{d\phi}{dt} \| \ dt \  = \  \int_0^1 \| \mathcal{H}_t \| \  dt  
\end{equation}
Here \  $\mathcal{H}_t$ \ the unique Hamiltonian \ $\mathcal{H}: [0,1] \times\mathcal{M} \rightarrow \mathbb{R}$ \ obeying 
\begin{equation}
              \int_\mathcal{M} \mathcal{H}_t  \ \omega^n \ = \ 0 
\end{equation}
for all \ $t\in [0,1]$. \  Then, a Finslerian distance function \ $\mathsf{dis}$ \ between two elements \ $\phi_1, \phi_2 \ \in \ \mathsf{Ham}(\mathcal{M}, \omega)$ \ 
associated to the norm \ $\| \cdot \|$ \  is defined as 
\begin{equation}
                 \mathsf{dis}(\phi_1, \phi_2) \ = \ \inf_{\psi} \  l(\psi) 
\end{equation} 
where \ $\psi$ \ is a Hamiltonian path connecting \ $\phi_1$ \ and \ $\phi_2$ \ and the infimum is taken over all such paths. It turns out that \ $\mathsf{dis}$ \ is non-negative, 
it is symmetric, it is non-degenerate and it satisfies the triangle inequality, therefore it is a genuine metric on \ $\mathsf{Ham}(\mathcal{M}, \omega)$. \ The existence of such a metric is 
non-trivial since \ $\mathsf{Ham}(\mathcal{M}, \omega)$ \ is an infinite dimensional group, hence no such metric is guaranteed to exist. \\

As in the finite-dimensional case, the group \ $\mathsf{Ham}(\mathcal{M}, \omega)$ \ acts on its Lie algebra \ 
$C_0^\infty(\mathcal{M})$ \  through the adjoint action       
\begin{equation}
           Ad_\phi f \ =  f \circ \phi^{-1}
\end{equation}
for \ $\phi \in \mathsf{Ham}(\mathcal{M}, \omega)$ \ and for \ $f\in C_0^\infty(\mathcal{M})$. \ One can see that the distance function (30) is bi-invariant under the adjoint action, 
namely
\begin{equation}
       \mathsf{dist}(\phi_1, \phi_2) \ = \ \mathsf{dist}(\phi_1\psi, \phi_2\psi) \ = \ \mathsf{dist}(\psi\phi_1, \psi\phi_2)
\end{equation}
Hofer \cite{Hofer} considered the \ $L^\infty$ \ norm on $C_0^\infty (\mathcal{M})$ \ which induces bi-invariant metric on \ $(\mathbb{R}^{2n}, \omega_0)$. \  
His construction was generalized by \cite{LMcD} for general symplectic manifolds. We may recall at this point that two distance functions \ $d_1, \ d_2$ \ in a general metric space are
equivalent if there is a positive constant \ $C>0$ \  such that 
\begin{equation}  
       \frac{1}{C}\ d_1 \  \leq \  d_2 \  \leq \ C \ d_1 
\end{equation}
With these definitions, in a series of works  \cite{EliashP, OW}  culminating with \cite{BuhO}, it was proved that for a closed symplectic manifold \ $(\mathcal{M}, \omega)$, \ any bi-invariant 
Finslerian metric on the set of its Hamiltonian diffeomorphisms induced by a norm  on \ $C_0^\infty (\mathcal{M})$ \  which is continuous with respect to the \ $C^\infty$ \ topology is either 
zero or equivalent to Hofer's metric. All other \ $L^s$ \ metrics are zero \cite{EliashP}. Hofer's metric, which belongs to the class \ $L^\infty$ \ is the unique non-trivial metric among all \ $L^s$ \ bi-invariant 
metrics under the group of Hamiltonian diffeomorphisms on a closed symplectic manifold.  \\

Incidentally, the approach of understanding the geometry of some space by understanding properties of its group of automorphisms/symmetries is what the ``Erlangen programme" proposed by F. Klein (1872) 
is all about. This is what we follow in this section to argue for the suitability of cubes for the phase space coarse-graining.\\

It might appear in this result that the metric equivalence (33), which is not an equality, would compromise the uniqueness of Hofer's metric, hence the exclusive use of its unit balls which are cubes in 
phase space coarse-graining. However this is not the case. To see this, we only have to recall  the forms of both the Boltzmann and the Gibbs expressions for the entropy. Let us use the latter for concreteness. 
We have, if we are careful, that the Gibbs entropy is defined by 
\begin{equation}     
       \mathcal{S}_G \ = \ -k_B \int_\mathcal{M} \ \rho(x) \ \log \left\{ \frac{\rho(x)}{\rho_0} \right\} \ d\mathsf{Vol}_\mathcal{M}  
\end{equation}
In this expression \ $\rho_0$ \ is taken to be a constant distribution on \ $(\mathcal{M}, \omega)$. \ It is present in order to make the logarithm dimensionless. The constant distribution \ $\rho_0$ \  
has no physical significance, since it is only differences/derivatives of the entropy that have any physical meaning, and not its absolute value. \\

A redefinition of distances by a constant factor \ $C$ \ as
indicated in (33), can be absorbed through a redefinition of \ $\rho_0$, \ in both the Boltzmann and the Gibbs entropy formulae.  As such,  the metric equivalence assumed in the proof of the uniqueness of 
Hofer's metric has no implications for the coarse-graining process of the classical phase space \ $(\mathcal{M}, \omega)$. \ 
The cubes which are the cells of the phase space coarse-graining process, are replaced by cubes of a different side length, under such an equivalence. 
Consequently, all the points made in this section remain valid even considering two equivalent (33), but not equal, metrics. \\

%%%%%%%%%%%%%%%%%%%%%%%%%%%%%%%%%%%%%%%%%%%%%%%%%%%%%%%%%%%%%%%%%%%%%%%%%%%%%%%%%%%

\section{Discussion and outlook}

We presented in this work some comments on possible implications of the symplectic non-squeezing theorem for Statistical Physics and for coarse-graining. 
It is a fact that very few things are really known about high-dimensional manifolds which would be the phase spaces of systems of many degrees of freedom. 
But, some things are known about infinite dimensional manifolds \cite{Kuk, AbbM} being the phase spaces of some nonlinear partial differential equations.  
The  latter may have some common features with the former, which may turn out to be useful not only for the thermodynamic limit of particle systems, 
but also for their implications to field theories, as long as relevant model-independent results can be obtained. 
Given that symplectic structures are inherently two-dimensional, and that symplectic manifolds lack any local geometry, it may be of some interest to see 
what new results, if any,  may  come out of  such investigations.  \\ 

On the other hand, more structure is present in a system of many degrees of freedom, such as a metric. 
One can probe several features of a system by following a metric approach. But such a choice of a metric is not unique. 
And a symplectic structure always induces sets of almost complex structures on a symplectic manifold \cite{McDS}.  
The problem is that such  almost complex structures, even if compatible or tame,   are usually not integrable.
If they were, then high dimensional phase spaces would be K\"{a}hler manifolds, which have a lot of structure and about which several things are known, 
since K\"{a}hler manifolds have symplectic, Hermitian and algebraic-geometric properties. But most symplectic manifolds are not K\"{a}hler, 
and this fact begs for a different approach \cite{G-Pseudo}.\\  

A slightly different  question, which may allow us to reach some conclusions pertinent to our goals, 
is  whether a high dimensional phase space of a physical system, is reasonably close 
(in the rather strong \  $C^\infty$, \ or in  the weaker measured Gromov-Hausdorff sense) to a K\"{a}hler manifold. 
If this is true, it might allow us to draw some conclusions about the underlying physical system.  
An immediate/uneducated guess to answering this question would be negative. 
If such an answer turns out to be true, one might narrow the scope of the question and ask  which systems have phase spaces for which 
such a statement might be true and what its implications might be.  We believe that it may be worth  further investigating this matter in the future.   \\  

The obvious weak point of the present treatment is the lack of concrete examples on which explicit calculations can be performed, 
and have such ideas checked for their physical relevance and significance.  This however has been a persistent problem plaguing the dynamical 
foundations of Statistical Mechanics for decades. Analytical results are scarce, and hard to establish from first principles,  
especially if one does not assume classical ensemble equivalence, is interested in far from equilibrium phenomena, 
explores the treatment of systems having long-range interactions, and even going as far as considering the use of different, from the Boltzmann and 
Gibbs, entropic functionals. Coming up with physically relevant but also analytically tractable models to test some of the above ideas is a challenge that is 
certainly worth pursuing, in our opinion. \\

%%%%%%%%%%%%%%%%%%%%%%%%%%%%%%%%%%%%%%%%%%%%%%%%%%%%%%%%%%%%%%%%%%%%%%%%%%%%%%%%%%%

                                 \vspace{5mm}

\noindent{\bf Acknowledgements:} The author is grateful to Prof. Panayotis Benetatos for numerous discussions on the topics of the present work, and of Statistical
Physics in general, for his continuous  encouragement to keep pursuing such foundational issues and for providing  many references on which this work relies. 
The author is also grateful to Prof. Anastasios Bountis for his  encouragement, and constant  support over many years, which made the present work possible. \\

%%%%%%%%%%%%%%%%%%%%%%%%%%%%%%%%%%%%%%%%%%%%%%%%%%%%%%%%%%%%%%%%%%%%%%%%%%%%%%%%%%%

%%%%%%%%%%%%%%%%%%%%%%%%%%%%%%%%%%%%%%%%%%%%%%%%%%%%%%%%%%%%%%%%%%%%%%%%%%%%%%%%%

\end{document}